\documentclass[12pt]{article}
\usepackage{subfloat}
\pdfoutput = 1
\usepackage{graphics}
\usepackage{graphicx} %Include figure filesusepackage{graphicx} %Include figure files
\begin{document}
\date{Today}
%%%%%%%%%%%%%%%%%%%%
\title{{\bf{\Large Constraints on rainbow gravity functions from black hole thermodynamics }}}
%%%%%%%%%%%%%%%%%%%%

\author{
{\bf {\normalsize Sunandan Gangopadhyay}$^{a,c}$
\thanks{sunandan.gangopadhyay@gmail.com, sunandan@associates.iucaa.in}},\,
{\bf {\normalsize Abhijit Dutta}$^{b, a}
$\thanks{dutta.abhijit87@gmail.com}}\\
$^{a}$ {\normalsize Department of Physics, West Bengal State University, Barasat, Kolkata 700126, India }\\[0.2cm]
$^{b}$ {\normalsize Department of Physics, Adamas University, Barasat, Kolkata 700126, India }\\[0.2cm]
$^{c}${\normalsize Visiting Associate in Inter University Centre for Astronomy \& Astrophysics (IUCAA),}\\
{\normalsize Pune, India}\\[0.1cm]
}
\date{}

\maketitle

\begin{abstract}
{\noindent In this paper, we investigate the thermodynamic properties of black holes in the framework of rainbow gravity. By considering rainbow functions in the metric of Schwarzschild and Reissner-Nordstr\"{o}m black holes, remnant and critical masses are found to exist.
Demanding the universality of logarithmic corrections to the semi-classical
area law for the entropy leads to constraining the form of the rainbow functions. The mass output and the radiation rate for these constrained form
of rainbow functions have been computed for different values of the rainbow parameter $\eta$ and have striking similarity to those derived from the generalized uncertainty principle. }
\end{abstract}
\vskip 1cm

%%%%%%%%%%%%%%%%%%%%%%%%%%%%%%%%%%%%%%%%%%%%%%%%%%%%%%%%%%%%%%%%

\section{Introduction}

There is a strong indication of an observer independent minimum length scale in all theories of quantum gravity, e.g. in string theory \cite{amati}, noncommutative geometry \cite{girelli}, loop quantum gravity 
\cite{rovelli, carlip} and Lorentzian dynamical triangulations 
\cite{ambjorn1}-\cite{ambjorn3} to name a few. This minimum measurable length scale is assumed to be the Planck scale. The mathematical apparatus of general theory theory of relativity is based on a smooth manifold. This picture breaks down when spacetime is probed at energies of the order of Planck energy \cite{Maggiore,Park}. One therefore expects a radically new
picture of spacetime. This includes departing from the standard
relativistic dispersion relation implying the breaking of Lorentz invariance.  
Lorentz symmetry which is well known to be one of the most remarkable symmetries in nature fixes the standard form of energy-momentum dispersion relation to be $E^2-p^2=m^2$. It has been suggested in most theories of quantum gravity \cite{Hooft}-\cite{Carroll} that in the ultraviolet limit, the standard energy-momentum dispersion relation gets modified.
For example, in Horava-Lifshitz gravity, the energy-momentum dispersion relation gets modified in the ultraviolet region \cite{horava1, horava2}. 
Due to the presence of a maximum energy scale $E_p$ leading eventually to the
breaking of Lorentz invariance, theories have been constructed with
two universal invariants, namely, the velocity of light $c$ and the Planck energy $E_p$. This leads to modified energy-momentum dispersion relations (MDR). 
This generalization of special relativity is called double special relativity (DSR) \cite{Magueijo-smolin1}-\cite{Magueijo-smolin3}.

In curved spacetime it is possible to make generalization of DSR leading to double general relativity \cite{amelino-camelia2}. In these theories the geometry of spacetime depends on the energy of the particle used to probe it.
Therefore the geometry here is represented by a one parameter family of
energy dependent metrics forming a rainbow of metrics. Hence the name rainbow gravity. The effects of rainbow gravity have led to considerable changes in the thermodynamic properties of black holes \cite{faiz1}-\cite{faiz3}. However, the
entropy expression does not contain the logarithmic corrections (which are well known to be universal) to the semi-classical area law for all forms of the rainbow functions.

In this paper we start by studying the modification of the thermodynamic properties, namely, the Hawking temperature, heat capacity and entropy for the rainbow inspired Schwarzschild and Reissner-Nordstr\"{o}m (RN) black holes. We compute the critical and remnant masses for these black holes below which the thermodynamic quantities become ill-defined. It is observed that the rainbow functions prevent the total evaporation of the black holes keeping a remnant
in exactly the same way as done by the generalized uncertainty principle (GUP) \cite{adler}-\cite{abhijit4}. We further observe that demanding the universality of logarithmic corrections to the semi-classical area law for entropy constraints the form of the rainbow functions. The expressions for entropy have a striking similarity to those derived using the GUP for these forms of the rainbow functions.
We then make use of the Stefan-Boltzmann law to estimate the mass output and radiation rate as a function of time taking into account the effects of the rainbow functions. These characterstics are also observed to be very similar to those obtained from the GUP \cite{adler}.

The paper is organized as follows. In section 2, we present the basics  of rainbow gravity. In section 3 we study the thermodynamics of Schwarzschild black holes taking into account the effects of rainbow gravity. In section 3.1
we obtain the mass output and radiation rate of rainbow gravity inspired Schwarzschild black holes as a function of time for different values of the rainbow parameter $\eta$. In section 4 we study the thermodynamics of Reissner-Nordstr\"{o}m black holes taking into account the effects of rainbow gravity. Finally, we conclude in section 5.

%%%%%%%%%%%%%%%%%%%%%%%%%%%%%%%%%%%%%%%%%%%%%%%%%%%%%%%%%%%

\section{Basics of rainbow gravity} 
Rainbow gravity generalizes the modified dispersion relations in DSR to curved spacetime. These modified dispersion relations are given by \cite{Magueijo-smolin1,Magueijo-smolin2}
\begin{eqnarray}
E^2f^2\left(E/E_p\right) - p^2 g^2\left(E/E_p\right) = m^2c^4
\end{eqnarray}
where $E_p$ is the Planck Energy and the functions $f\left(E/E_p\right)$ and  $g\left(E/E_p\right)$ are called rainbow functions. These functions are responsible for the modification of the energy-momentum relation in the ultraviolet limit. However, they should be constrained to reproduce standard dispersion relation in the infrared limit. This requires
\begin{eqnarray}
 \lim\limits_{E/E_p\rightarrow 0}f\left(E/E_p\right) = 1 ~; \lim\limits_{E/E_p\rightarrow 0}g\left(E/E_p\right) = 1 ~.
\label{rafu}
\end{eqnarray}
In terms of a one-parameter family of orthonormal frame fields, a one-parameter family of energy-dependent metrics can be expressed as \cite{Magueijo-smolin3}
\begin{eqnarray}
g = \eta^{ab}e_a\left(E/E_p\right)e_b\left(E/E_p\right)
\label{rafu1}
\end{eqnarray}
where 
\begin{eqnarray}
e_0\left(E/E_p\right) = \frac{1}{f\left(E/E_p\right)}\tilde{e}_0\\  e_i\left(E/E_p\right) = \frac{1}{g\left(E/E_p\right)}\tilde{e}_i~.
\label{rafu2}
\end{eqnarray}
The quantities with a tilde stand for energy independent frame fields.
In the limit $E/E_p \rightarrow 0$, usual general relativity is recovered. The field equations of Einstein are also modified accordingly
\begin{eqnarray}
G_{\mu \nu}\left(E/E_p\right) = 8\pi G\left(E/E_p\right)T_{\mu \nu}\left(E/E_p\right)
\label{modified Einstein field equations}
\end{eqnarray}
where the energy dependent Newton's universal gravitational constant 
$G\left(E/E_p\right)$ becomes the conventional Newton's universal gravitational constant $G = G(0)$ in the limit  $E/E_p\rightarrow0$. Due to this modification the corresponding black hole metrics also get redefined. In this work we shall choose specific rainbow functions \cite{Amelino-Camelia1} which are motivated from loop quantum gravity considerations \cite{alfaro}-\cite{smolin} 
\begin{eqnarray}
f(E/E_p) = 1~ ;~ g(E/E_p) = \sqrt{1-\eta \left(\frac{E}{E_p}\right)^n}
\label{rainbow functions}
\end{eqnarray}
where $\eta$ is the rainbow parameter. In the subsequent discussion, we shall use natural units $c=1=\hbar$ and $k_B=1$.

%%%%%%%%%%%%%%%%%%%%%%%%%%%%%%%%%%%%%%%%%%%%%%%%%%%%%%%%%%
%%%%%%%%%%%%%%%%%%%%%%%%%%%%%%%%%%%%%%%%%%%%%%%%%%%%%%%%%% 
 
\section{Thermodynamics of rainbow gravity inspired Schwarzschild black holes}
In this section we want to study the thermodynamic properties of Schwarzschild black holes taking into account the effect of the rainbow functions (\ref{rainbow functions}). The metric of the Schwarzschild black hole inspired by rainbow gravity is given by \cite{Magueijo-smolin3}
\begin{eqnarray}
ds^2 = -\frac{1}{f^2(E/E_p)}\left(1-\frac{2MG}{r}\right)dt^2+\frac{1}{g^2(E/E_p)}\left(1-\frac{2MG}{r}\right)^{-1}dr^2+\frac{r^2}{g^2(E/E_p)}d\Omega^2.
\label{Modified Schwarzschild metric }
\end{eqnarray}
The surface gravity is related with the Hawking temperature as \cite{hawking} \begin{eqnarray}
T = \frac{\kappa}{2\pi}
\label{relation between HT and SG}
\end{eqnarray}
where the surface gravity is defined by the relation
\begin{eqnarray}
\kappa = \lim_{r\to\ R_s}\sqrt{-\frac{1}{4}g^{rr}g^{tt}\left(g_{tt,r}\right)^2}
\label{surface gravity}
\end{eqnarray}
where $R_s = 2GM$ is the Schwarzschild radius. 
Hence, the surface gravity for the rainbow gravity inspired Schwarzschild black hole reads
\begin{eqnarray}
\kappa=\frac{g(E/E_p)}{f(E/E_p)}\frac{1}{4MG}~.
\label{Surface gravity}
\end{eqnarray}
Therefore, the Hawking temperature is given by
\begin{eqnarray}
T =\frac{1}{8\pi G}\sqrt{\frac{1}{M^2}-\frac{\eta}{\left(2GE_p\right)^n}\frac{1}{M^{n+2}}}~.
\label{Modified Hawking Temperature}
\end{eqnarray}
In obtaining the above expression we have set $E=\frac{1}{2GM}$.
This equation gives a relation between the temperature and mass of the rainbow gravity inspired Schwarzschild black hole.
Since the temperature has to be a real quantity, we obtain the following condition  
\begin{eqnarray}
\frac{1}{M^2}-\frac{\eta}{\left(2GE_p\right)^n}\frac{1}{M^{n+2}}\geq 0.
\label{real_Temp_Cond}
\end{eqnarray}
The above condition readily leads to the existence of a critical mass below which the temperature becomes a complex quantity  
\begin{eqnarray}
M_{cr} =  \frac{{\eta}^{\frac{1}{n}}}{2GE_p}=\eta^{\frac{1}{n}}M_p~.
\label{critical mass}
\end{eqnarray}
The heat capacity of the rainbow gravity inspired Schwarzschild black hole reads
\begin{eqnarray}
C = \frac{dM}{dT} = \frac{16\pi G\sqrt{\frac{1}{M^2}-\frac{\eta}{\left(2GE_p\right)^n}\frac{1}{M^{n+2}}}}{\frac{(n+2)\eta}{\left(2GE_p\right)^n}\frac{1}{M^{n+3}}-\frac{2}{M^3}}~.
\label{heat capacity}
\end{eqnarray}
The remnant mass (where the black hole stops evaporating) can be obtained by setting $C = 0$. This yields 
\begin{eqnarray}
M_{rem} = \frac{{\eta}^{\frac{1}{n}}}{2GE_p}=\eta^{\frac{1}{n}}M_p~.
\label{remnant mass}
\end{eqnarray}
Thus, we have demonstrated that the remnant mass of the black hole is equal to its critical mass. 

\noindent The entropy can be calculated using the heat capacity of this black hole as follows 
\begin{eqnarray}
S = \int{C\frac{dT}{T}} =  \int{\frac{dM}{T}}~.
\label{entropy1}
\end{eqnarray}
\noindent Substituting eq.(\ref{Modified Hawking Temperature}) in eq.(\ref{entropy1}) and carrying out a binomial expansion keeping 
terms upto  $\mathcal{O}(\eta^3)$ and assuming that $n\geq 3$ leads to
\begin{eqnarray}
S &=& 8\pi G\int{\frac{dM}{\sqrt{\frac{1}{M^2}-\frac{\eta}{\left(2GE_p\right)^n}\frac{1}{M^{n+2}}}}}\nonumber\\
 &=&  8\pi G \int{\left[M+\frac{ \eta}{2\left(2GE_p\right)^n}\frac{1}{M^{n-1}}+\frac{3 {\eta}^2}{8\left(2GE_p\right)^{2n}}\frac{1}{M^{2n-1}}+\frac{5 {\eta}^3}{16\left(2GE_p\right)^{3n}}\frac{1}{M^{3n-1}}\right]dM}\nonumber\\
&=& 8\pi G\left[\frac{M^2}{2}+\frac{ \eta}{2(2-n)\left(2GE_p\right)^n}\frac{1}{M^{n-2}}+\frac{{3\eta}^2}{8(2-2n)\left(2GE_p\right)^{2n}}\frac{1}{M^{2n-2}}+\frac{{5\eta}^3}{16(2-3n)\left(2GE_p\right)^{3n}}\frac{1}{M^{3n-2}}\right]\nonumber\\
&=& S_{BH}+\frac{{\pi}^{\frac{n}{2}} \eta}{(2-n)}\frac{1}{{S_{BH}}^{(\frac{n}{2}-1)}}+\frac{3{\pi}^n{\eta}^2}{8(1-n)}\frac{1}{{S_{BH}}^{(n-1)}}+\frac{5{\pi}^{\frac{3n}{2}} {\eta}^3}{8(2-3n)}\frac{1}{{S_{BH}}^{(\frac{3n}{2}-1)}}
\label{gen_Entropy_Expn}
\end{eqnarray}
where $S_{BH}=\frac{4\pi M^2}{M_p^2}$ is the semi-classical Bekenstein-Hawking entropy for the Schwarzschild
black hole. The reason for assuming $n \geq 3$ is that the result of the integration is not valid for $n = 1,2$ as can be seen easily from the integrand. In terms of the area of the horizon $A=4\pi r_{s}^2 =16\pi G^2 M^2 = 4 l_p^2 {S_{BH}}$, the above expression for the entropy can be recast in the form
\begin{eqnarray}
S = \frac{A}{4}+\frac{{\pi}^{\frac{n}{2}} \eta}{(2-n)}\frac{1}{{\left(\frac{A}{4}\right)}^{(\frac{n}{2}-1)}}+\frac{3{\pi}^n{\eta}^2}{8(1-n)}\frac{1}{{\left(\frac{A}{4}\right)}^{(n-1)}}+\frac{5{\pi}^{\frac{3n}{2}} {\eta}^3}{8(2-3n)}\frac{1}{{\left(\frac{A}{4}\right)}^{(\frac{3n}{2}-1)}}
\label{gen. entropy in terms of area}
\end{eqnarray}
where we have set $l_p =1$.
It is evident from this expression for the entropy of the black hole that there are no logarithmic corrections to the semi-classical result for the values of $n \geq 3$. However if one takes into account the universality of the logarithmic corrections then the values of $n$ gets restricted to $n = 1,2$.

\noindent For $n=1$, the entropy expression upto $\mathcal{O}(\eta^2)$ in terms of horizon area takes the form
\begin{eqnarray}
S &=&  8\pi G \int{\left[M+\frac{ \eta}{4GE_p}+\frac{3 {\eta}^2}{8\left(2GE_p\right)^{2}}\frac{1}{M}\right]dM}\nonumber\\
&=& S_{BH}+\eta\sqrt{\pi}\sqrt{S_{BH}}+\frac{3{\pi}{\eta}^2}{8}\ln \left(S_{BH}\right)+\frac{3{\pi}{\eta}^2}{8}\ln{4\pi}  
\nonumber\\
&=& \frac{A}{4}+\eta\sqrt{\pi}\sqrt{\frac{A}{4}}+\frac{3{\pi}{\eta}^2}{8}\ln \left(\frac{A}{4}\right)+\frac{3{\pi}{\eta}^2}{8}\ln({4\pi}).
\label{Entropy for n=1}
\end{eqnarray}
The above expression involves the logarithmic corrections which are known to be universal. Remarkably, the expression for the entropy has a striking similarity with the expression for entropy obtained using generalized uncertainty principle with a linear term in the momentum uncertainty \cite{abhijit2}. This is evident from the presence of the $\sqrt{A}$ term in the entropy expression.

\noindent For $n=2$, the entropy expression becomes
\begin{eqnarray}
S = \frac{A}{4}+\frac{\eta \pi}{2}\ln\left(\frac{A}{4}\right)+\frac{\eta \pi}{2}\ln\left(4\pi\right)-\frac{3{\pi}^2{\eta}^2}{8}\frac{1}{{\left(\frac{A}{4}\right)}}~.
\label{Entropy-n2}
\end{eqnarray}
This expression is similar in structure with the one obtained from the simplest generalized uncertainty principle (without any linear term in momentum uncertainty) \cite{adler}, \cite{abhijit1}.
%%%%%%%%%%%%%%%%%%%%%%%%%%%%%%%%%%%%%%%%%%%%%%%%%%%%%%%%%%%%%%%%%%%%%%%%%%%%%

\subsection{Energy output as a function of time }

If the temperature of the black hole is greater than the ambient temperature, then it must radiate energy in terms of photons and other ordinary particles. As a consequence of this, the mass of the black hole  reduces further while its temperature keeps on increasing. If one assumes that the energy loss is dominated by photons, then the Stefan-Boltzmann law can be employed to estimate the energy radiated as a function of time. For the standard case (in the limit $\eta\rightarrow0$) \cite{adler}, we have
\begin{eqnarray}
\frac{d}{dt}\left(\frac{M}{M_p}\right) = -\frac{\sigma}{256\pi^3G^2M_p^3}\left(\frac{M_p}{M}\right)^2
\label{em1}
\end{eqnarray}
where $\sigma$ is the Stefan-Boltzmann constant.
Setting $x = \frac{M}{M_p}$ and defining the characteristic time $t_{ch} = \frac{256\pi^3G^2M_p^3}{\sigma}$, the above equation takes the form
\begin{eqnarray}
\frac{dx}{dt} = -\frac{1}{t_{ch}x^2}~.
\label{eqn}
\end{eqnarray}
If $x_i$ refers to the initial mass at time $t=0$, the solution of 
this equation yields the mass-time relation to be
\begin{eqnarray}
x = \left(x_i^3 - \frac{3t}{t_{ch}}\right)^\frac{1}{3}~.
\label{Stefan law for standard case}
\end{eqnarray}
Hence the rate at which energy is radiated as a 
function of time takes the form 
\begin{eqnarray}
\frac{dx}{dt} = -\frac{1}{t_{ch}\left(x_i^3 - \frac{3t}{t_{ch}}\right)^\frac{2}{3}}~.
\end{eqnarray}
Eq.(\ref{Stefan law for standard case}) implies that the black hole evaporates completely in time $\frac{t}{t_{ch}} = \frac{1}{3}\left(\frac{M_i}{M_p}\right)^3$ and the rate at which energy is radiated blows up at the end of the process. 

We now proceed to carry out the above analysis with the rainbow gravity inspired Schwarzschild black hole. In the subsequent discussion, we shall restrict the values of $n$ in the rainbow functions to be $1,2$.

\noindent For $n=1$, the differential equation expressing the rate at which energy is radiated takes the form (in terms of $x$)
\begin{eqnarray}
\frac{dx}{dt} = -\frac{1}{t_{ch}x^2}\left(1-\frac{\eta}{x}\right)^2~.
\label{eqn-d}
\end{eqnarray}
Solving this, we obtain the mass-time relation upto $\mathcal{O}(\eta)$ to be
\begin{eqnarray}
x = \left[-\frac{3t}{t_{ch}} + x_i^3 + 3\eta x_i^2 -3\eta\left(x_i^3 - \frac{3t}{t_{ch}}\right)^\frac{2}{3}\right]^\frac{1}{3}~.
\label{zz}
\end{eqnarray}
Hence the rate at which energy is radiated as a function of time is given by
\begin{eqnarray}
\frac{dx}{dt} = -\frac{1}{t_{ch}\left[-\frac{3t}{t_{ch}} + x_i^3 + 3\eta x_i^2 -3\eta\left(x_i^3 - \frac{3t}{t_{ch}}\right)^\frac{2}{3}\right]^\frac{2}{3}}\left[1-\frac{\eta}{\left[-\frac{3t}{t_{ch}} + x_i^3 + 3\eta x_i^2 -3\eta\left(x_i^3 - \frac{3t}{t_{ch}}\right)^\frac{2}{3}\right]^\frac{1}{3}}\right]^2.
\label{energy output for n=1}
\end{eqnarray}
The time in which the black hole evaporates completely is given by (upto $\mathcal{O}(\eta)$)
\begin{eqnarray}
\frac{t}{t_{ch}}=\frac{x_{i}^3}{3}+\eta x_{i}^2~.
\label{eqn-d-time}
\end{eqnarray}
For $n = 2$, the differential equation expressing the rate at which energy is radiated takes the form (in terms of $x$)
\begin{eqnarray}
\frac{dx}{dt} = -\frac{1}{t_{ch}x^2}\left(1-\frac{\eta}{x^2}\right)^2~.
\label{eqn-df}
\end{eqnarray}
Solving this, we obtain the mass-time relation upto $\mathcal{O}(\eta)$ to be
\begin{eqnarray}
x = \left[-\frac{3t}{t_{ch}} + x_i^3 + 6\eta x_i -6\eta\left(x_i^3 - \frac{3t}{t_{ch}}\right)^\frac{1}{3}\right]^\frac{1}{3}~.
\label{zzz}
\end{eqnarray}
Hence the rate at which energy is radiated as a function of time is given by
\begin{eqnarray}
\frac{dx}{dt} = -\frac{1}{t_{ch}\left[-\frac{3t}{t_{ch}} + x_i^3 + 6\eta x_i -6\eta\left(x_i^3 - \frac{3t}{t_{ch}}\right)^\frac{1}{3}\right]^\frac{2}{3}}\left[1-\frac{2\eta}{\left[-\frac{3t}{t_{ch}} + x_i^3 + 6\eta x_i -6\eta\left(x_i^3 - \frac{3t}{t_{ch}}\right)^\frac{1}{3}\right]^\frac{2}{3}}\right].
\label{energy output for n=2}
\end{eqnarray}
The time in which the black hole evaporates completely is given by (upto $\mathcal{O}(\eta)$)
\begin{eqnarray}
\frac{t}{t_{ch}}=\frac{x_{i}^3}{3}+2\eta x_{i}~.
\label{d-time}
\end{eqnarray}
The mass as a function of time given by eq.(s)(\ref{zz}, \ref{zzz}) and the rate at which energy is radiated as a function of time  given by eq.(s)(\ref{energy output for n=1}, \ref{energy output for n=2}) are shown in Figures 1,2,3,4 for different values of $\eta$.
\begin{figure}[H]
\centering
\begin{minipage}{.45\linewidth}
  \includegraphics[width=\linewidth]{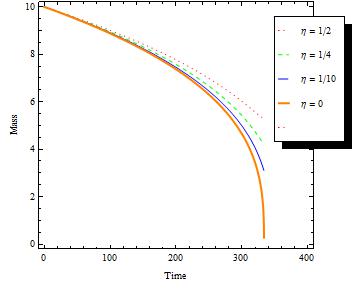}
  \caption{The mass of the black hole vs time for n=1. The mass is in unit of Planck mass and the time is in units of characteristic time. }
\end{minipage}
\hspace{.08\linewidth}
\begin{minipage}{.45\linewidth}
\includegraphics[width=\linewidth]{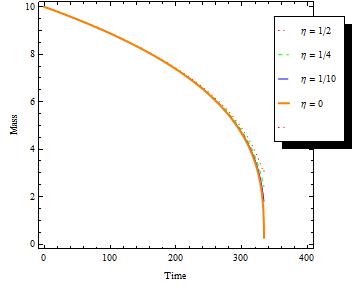}
\caption{The mass of the black hole vs time for n=2. The mass is in unit of Planck mass and the time is in units of characteristic time.}
\end{minipage}
\end{figure}

\noindent From the plots we observe that they are similar in appearance  with the plots which takes into account the effect of the generalized uncertainty principle in black hole thermodynamics \cite{adler}.

~~~~~~~~~~~~~~~~~~~~~~~~~~~~~~~~~~~~~~~~
~~~~~~~~~~~~~~~~~~~~~~~~~~~~~~~~~~~~~~~~
~~~~~~~~~~~~~~~~~~~~~~~~~~~~~~~~~~~~~~~
\begin{figure}[H]
\centering
\begin{minipage}{.45\linewidth}
  \includegraphics[width=\linewidth]{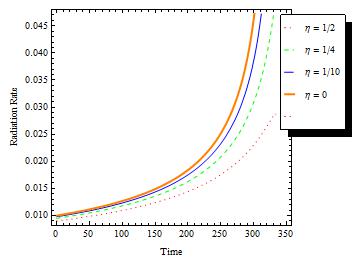}
  \caption{The radiation rate vs time for n=1. The rate is in unit of Planck mass per characteristic time and the time is in units of characteristic time. }
\end{minipage}
\hspace{.08\linewidth}
\begin{minipage}{.45\linewidth}
  \includegraphics[width=\linewidth]{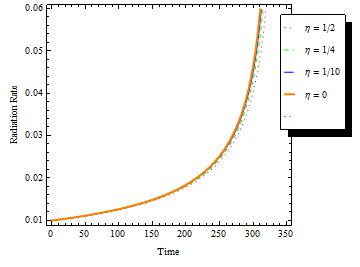}
  \caption{The mass of the black hole vs time for n=2. The mass is in unit of Planck mass and the time is in units of characteristic time.}
\end{minipage}
\end{figure}

~~~~~~~~~~~~~~~~~~~~~~~~~~~~~~~~~~~~~~~~
~~~~~~~~~~~~~~~~~~~~~~~~~~~~~~~~~~~~~~~~
~~~~~~~~~~~~~~~~~~~~~~~~~~~~~~~~~~~~~~~

\section{Thermodynamics of rainbow gravity inspired Reissner-Nordstr\"{o}m black holes  }

The rainbow gravity inspired Reissner-Nordstr\"{o}m black hole metric reads
\begin{eqnarray}
ds^2 = -\frac{1}{f(E/E_p)^2}\left(1-\frac{2M}{r}+\frac{Q^2}{r^2}\right)dt^2+\frac{1}{g(E/E_p)^2}\left(1-\frac{2M}{r}+\frac{Q^2}{r^2}\right)^{-1}dr^2+\frac{r^2}{g(E/E_p)^2}d\Omega^2~.
\label{R-N metric }
\end{eqnarray}
The surface gravity in this case is given by
\begin{eqnarray}
\kappa
=\frac{g(E/E_p)}{f(E/E_p)}\left(\frac{M}{r_0^2}-\frac{Q^2}{r_0^3}\right)
\label{R-N Surface gravity}
\end{eqnarray}
where $r_0 = M+\sqrt{M^2-Q^2}$ is the radius of the event horizon of the rainbow 
gravity inspired RN black hole.
The Hawking temperature therefore reads
\begin{eqnarray}
T = \frac{1}{2\pi}\sqrt{1-\frac{\eta}{\left(E_p\right)^n}\frac{1}{r_0^n}}\left(\frac{M}{r_0^2}-\frac{Q^2}{r_0^3}\right).
\label{R-N Modified Hawking Temperature}
\end{eqnarray}
Condition for the reality of the temperature yields
\begin{eqnarray}
1-\frac{\eta}{\left(E_p\right)^n}\frac{1}{r_0^n}\geq 0
\label{R-N real_Temp_Cond}
\end{eqnarray}
and this in turn implies that the critical mass of the rainbow gravity inspired RN black hole is given by
\begin{eqnarray}
M_{cr} =  \sqrt{\frac{Q^2}{2}+\frac{Q^4}{4\left(\frac{\eta}{E_p^n}\right)^{2/n}}+\frac{\left(\frac{\eta}{E_p^n}\right)^{2/n}}{4}}
\label{R-N critical mass}
\end{eqnarray}
The heat capacity of this black hole reads
 \begin{eqnarray}
 C = 4\pi \frac{(r_0-M)\sqrt{1-\frac{\eta}{\left(E_p r_o\right)^n}}}{\left(-\frac{2M}{r_0^2}+\frac{3Q^2}{r_0^3}\right)\left(1-\frac{\eta}{\left(E_p r_o\right)^n}\right)+\frac{n \eta }{E_p^n r_0^n}\left(\frac{M}{r_0^2}-\frac{Q^2}{r_0^3}\right)}~.
\label{R_N Heat Capacity}
\end{eqnarray}
As before, the remnant mass is obtained by setting $C=0$ :
\begin{eqnarray}
M_{rem} = \sqrt{\frac{Q^2}{2}+\frac{Q^4}{4\left(\frac{\eta}{E_p^n}\right)^{2/n}}+\frac{\left(\frac{\eta}{E_p^n}\right)^{2/n}}{4}}
\label{R-N remnant mass}
\end{eqnarray}
and is found to be identical to the critical mass.

\noindent The entropy of this black hole is now computed keeping terms upto  $\mathcal{O}(\eta^3)$ 
\begin{eqnarray}
S &=& 2\pi \int{\frac{dM}{\sqrt{1-\frac{\eta}{E_p^n r_0^n}}\left(\frac{M}{r_0^2}-\frac{Q^2}{r_0^3}\right)}}\nonumber\\
&=& S_{BH}+\frac{{\pi}^{\frac{n}{2}} \eta}{(2-n)}\frac{1}{{S_{BH}}^{(\frac{n}{2}-1)}}+\frac{3{\pi}^n{\eta}^2}{8(1-n)}\frac{1}{{S_{BH}}^{(n-1)}}+\frac{5{\pi}^{\frac{3n}{2}} {\eta}^3}{8(2-3n)}\frac{1}{{S_{BH}}^{(\frac{3n}{2}-1)}}
\label{R-N gen_Entropy_Expn}
\end{eqnarray}
where $S_{BH}=\pi r_0^2$ is the semi-classical Bekenstein-Hawking entropy for the rainbow gravity inspired RN black hole. Note that the above integration is valid only for $n\geq 3$. Once again no logarithmic corrections are present in the expression for the entropy. However carrying out the integration for $n = 1,2$ and making use of the identity
\begin{eqnarray}
\frac{r_{0}}{(Mr_0 -Q^2)}=\frac{1}{(r_0 -M)}
\end{eqnarray}
\noindent one gets for $n=1$ 
\begin{eqnarray}
S=\frac{A}{4}+\eta\sqrt{\pi}\sqrt{\frac{A}{4}}+\frac{3{\pi}{\eta}^2}{8}\ln \left(\frac{A}{4}\right)
\label{R-N Entropy for n=1}
\end{eqnarray}
and for $n=2$ 
\begin{eqnarray}
S = \frac{A}{4}+\frac{\eta \pi}{2}\ln\left(\frac{A}{4}\right)-\frac{3{\pi}^2{\eta}^2}{8}\frac{1}{{\left(\frac{A}{4}\right)}}~.
\end{eqnarray}
Once again the expression for entropy for $n=1$ is similar in structure to that obtained using the GUP with a linear term in the momentum uncertainty \cite{abhijit2}. For $n=1$ the expression is similar in structure to that obtained using the GUP without a linear term in the momentum uncertainty \cite{adler}, \cite{abhijit1}.

%%%%%%%%%%%%%%%%%%%%%%%%%%%%%%%%%%%%%%%%%%%%%%%%%%%%%%%%%%%%%

%%%%%%%%%%%%%%%%%%%%%%%%%%%%%%%%%%%%%%%%%%%%%%%%%%%%%%%%%%%%%

\section{Conclusions}
\noindent In this paper we have investigated the modifications of the  thermodynamic properties of Schwarzschild and Reissner-Nordstr\"{o}m black holes taking into account the effects of rainbow gravity functions. We found that  rainbow gravity modifies the mass-temperature relationship and leads to the existence of a black hole remnant. We further observe that as in the case of the generalized uncertainty principle, here also the remnant and critical masses are identical. The computation of the entropy does not contain the universal logarithmic corrections for all forms of the rainbow functions. It is only for the values of $n=1,2$ appearing in the rainbow functions that the logarithmic corrections to the semi-classical area law for the entropy are found to exist.
The mass output and the radiation rate are finally computed for these constrained forms of the rainbow functions for different values of the rainbow parameter $\eta$ and have striking similarity to those derived from the generalized uncertainty principle. 

%%%%%%%%%%%%%%%%%%%%%%%%%%%%%%%%%%%%%%%%%%%%%%%%%%%%%%%%%%%%%%%%%%%%

\section{Acknowledement}

S.G. acknowledges the support by DST SERB under Start Up Research Grant (Young Scientist), File No.YSS/2014/000180.

%%%%%%%%%%%%%%%%%%%%%%%%%%%%%%%%%%%%%%%%%%%%%%%%%%%%%%%%%%%%%%%%%%%%

\end{document}